\documentclass[aps,prl,superscriptaddress,amsfonts,amsmath,amssymb,reprint,floatfix]{revtex4-1}
\usepackage{url}
\usepackage{bm}
\usepackage{graphicx}
\usepackage{amsmath}
\usepackage{amstext}
\usepackage{amssymb}
\usepackage{amsfonts}
\usepackage{amsbsy}
\usepackage{verbatim}
\usepackage{color}
\usepackage[colorlinks=true, urlcolor=blue, linkcolor=blue, citecolor=blue, pdftex]{hyperref}
\usepackage{multirow}

\AtBeginDocument{\usepackage{booktabs}}

\begin{document}

{\bf Comment on ``$Z_{2}$ spin liquid phase on the kagome lattice: a new saddle point'', by Tao Li 
[\href{http://arxiv.org/abs/1601.02165}{arXiv:1601.02165 (2016)}]}\\

In a recent paper, Tao Li~\cite{li2016} claimed that a gapped $\mathbb{Z}_2$ spin liquid, obtained using projected
Gutzwiller fermionic wave functions, can be stabilized in the Heisenberg model on the kagome lattice. According to
his calculations, the best wave function is gauge equivalent to the so-called $\mathbb{Z}_{2}[0,\pi]\beta$ state that has 
been proposed in Ref.~\cite{lu2011} and numerically studied by us~\cite{iqbal2011}. The major claim was that 
the $\mathbb{Z}_2$ state has a (slightly) lower energy than its ``parent'' $U(1)$ Dirac spin liquid, which is recovered once 
the spinon pairing is switched off. Calculations were done for the Heisenberg model with nearest-neighbor (NN) $J_1$ 
and next-nearest-neigbor (NNN) $J_2$ super-exchange couplings. The energy gain was claimed to be ``substantial'' for 
$J_2/J_1=0.15$, but also finite for $J_2=0$. However, the values of the energy gain were reported only in few cases, 
without performing a size scaling (numbers for $J_2=0$ were not reported), while the size scaling for some variational
parameters have been shown for $J_2/J_1=0.15$. 

In this comment, we perform very accurate variational calculations with both $U(1)$ and $\mathbb{Z}_2$ wave functions, with
much smaller error bars with respect to our recent calculations~\cite{iqbal2015}, and show that even though a small 
energy gain is found (compatible with our previous results), it goes to zero when increasing the size of the cluster, as already claimed in our previous work~\cite{iqbal2015}. 
We thus confirm with unprecedented accuracy the fact, that the $\mathbb{Z}_{2}$ spin liquid is a local energy minimum that 
goes away with system size. Therefore, our calculations confirm once more that the $U(1)$ spin liquid with Dirac 
nodes is stable against the opening of a spin gap, not only for $J_2=0$, but also for small values of $J_2/J_1$, e.g., 
$J_2/J_1=0.15$. 

The variational wave function for the spin model is obtained by applying the Gutzwiller projector ${\cal P}_{\rm G}$
to an uncorrelated wave function:
\begin{equation}
|\Psi_{\rm var}\rangle = {\cal P}_{\rm G}|\Phi_0\rangle,
\end{equation}
where $|\Phi_0\rangle$ is the ground state of a BCS Hamiltonian that contains NN real hopping ($\chi_1$, which can be
taken as the energy scale, i.e., $\chi_1=1$), NNN real hopping ($\chi_2$) and spinon pairing ($\Delta_2$), and two 
on-site terms, one for the chemical potential ($\mu$) and the other for the real on-site pairing ($\zeta_{\rm R}$). 
A non-trivial sign structure for the hopping and pairing terms is assumed~\cite{lu2011,iqbal2011}, e.g., a fictitious 
gauge magnetic flux is piercing each unit cell. The $U(1)$ Dirac state is obtained by taking $\Delta_2=0$ and 
$\zeta_{\rm R}=0$ (in this case the chemical potential is irrelevant, not changing the single-particle orbitals).

\begin{figure}
\vspace{0.1cm}
\includegraphics[width=0.99\columnwidth]{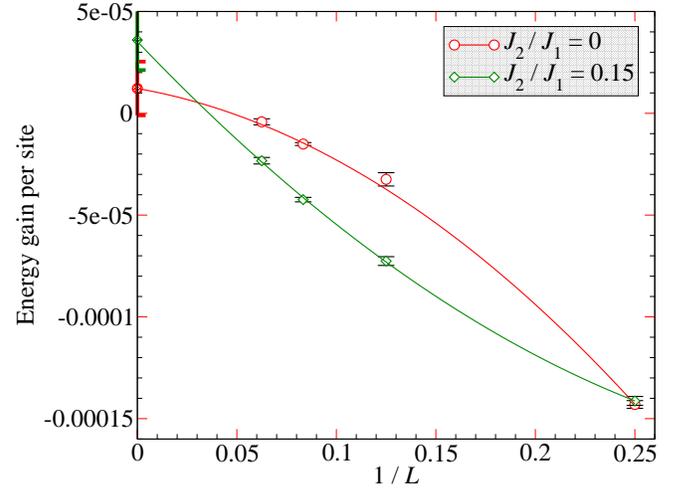}
\caption{The size scaling of the energy gain per site $\Delta E=E_{\mathbb{Z}_2}-E_{U(1)}$ is shown for $J_2=0$ and $J_2/J_1=0.15$, 
for $L=4$, $8$, $12$ and $16$ clusters. The results for $L=4$ and $8$ are from Ref.~\cite{iqbal2015}.}
\label{fig:fig1}
\end{figure}

Our results are summarized in Table~\ref{tab:energies} and Fig.~\ref{fig:fig1}. Here, we optimized $\chi_2$ for
the $U(1)$ state, while we used the variational parameters obtained in Refs.~\cite{li2016,privcomm} for the $\mathbb{Z}_2$ 
wave function and computed the energies of the gapless $U(1)$ and gapped $\mathbb{Z}_2$ states. Although the gapped spin 
liquid has a slighly lower energy than the gapless one, namely $\Delta E=-0.0000044(13)$ for $J_2=0$ and 
$-0.0000226(14)$ for $J_2/J_1=0.15$ on the $768$-site cluster, the size scaling clearly show that, in  the
thermodynamic limit, the $\mathbb{Z}_2$ {\it Ansatz} does not give a finite energy gain. Indeed, for $J_2/J_1=0$, the gain is 
zero within the errorbars, and for $J_2/J_1=0.15$, it is zero within two errorbars. Hence, the fact that the energy 
gain of the gapped $\mathbb{Z}_{2}[0,\pi]\beta$ spin liquid goes to zero in the thermodynamic limit is {\it irrefutable}. \\

\noindent Yasir Iqbal, \\
  Institute for Theoretical Physics and Astrophysics, \\
  Julius-Maximilian's University of W{\"u}rzburg,\\
  D-97074 W{\"u}rzburg, Germany

\vspace{0.2cm} 

\noindent Didier Poilblanc, \\
  Laboratoire de Physique Th\'eorique, \\
  CNRS and Universit\'e de Toulouse, F-31062 France

\vspace{0.2cm} 

\noindent Federico Becca, \\
  Democritos National Simulation Center, IOM and \\
  Scuola Internazionale Superiore di Studi Avanzati, \\
  I-34136 Trieste, Italy

\begin{table*}
\begin{tabular}{lllllll}
\hline \hline
 & \multicolumn{3}{c}{$12\times12\times3=432$-site cluster} & \multicolumn{3}{c}{$16\times16\times3=768$-site cluster} \\

\cmidrule(lr){2-4} \cmidrule(lr){5-7}
 
\multicolumn{1}{c}{$J_2/J_1$} & \multicolumn{1}{c}{$E_{U(1)}$} & \multicolumn{1}{c}{$E_{\mathbb{Z}_2}$} & \multicolumn{1}{c}{$\Delta E$} & \multicolumn{1}{c}{$E_{U(1)}$} & \multicolumn{1}{c}{$E_{\mathbb{Z}_2}$} & \multicolumn{1}{c}{$\Delta E$}   \\ 
\hline       
                                                                                    
\multirow{1}{*}{$0$}    & $-0.4287114(3)$ & $-0.4287266(4)$ & $-0.0000151(7)$  & $-0.4287160(6)$ & $-0.4287204(7)$ & $-0.0000044(13)$ \\ 
                                                      
\multirow{1}{*}{$0.15$} & $-0.4346363(4)$ & $-0.4346787(6)$ & $-0.0000424(11)$ & $-0.4347006(7)$ & $-0.4347232(7)$ & $-0.0000226(14)$ \\ 
\hline \hline

\end{tabular}
\caption{Ground state energies per site (in units of $J_1$) of the $U(1)$ Dirac state (current work) and the $\mathbb{Z}_{2}[0,\pi]\beta$
state (Refs.~\cite{li2016,privcomm}) together with the energy gain per site ($\Delta E=E_{\mathbb{Z}_{2}}-E_{U(1)}$). The 
corresponding parameters are given in Table~\ref{tab:parameters}.}
\label{tab:energies}
\end{table*}

\begin{table*}
\begin{tabular}{crrrrr}
\hline \hline
 & \multicolumn{5}{c}{$12\times12\times3=432$-site cluster} \\

\cmidrule(lr){2-6}
 
 & \multicolumn{1}{c}{$U(1)$} & \multicolumn{4}{c}{$\mathbb{Z}_{2}$} \\

\cmidrule(lr){2-2} \cmidrule(lr){3-6}

\multicolumn{1}{c}{$J_{2}/J_{1}$} & \multicolumn{1}{c}{$\chi_{2}$} & \multicolumn{1}{c}{$\Delta_{2}$} & \multicolumn{1}{c}{$\chi_{2}$} & \multicolumn{1}{c}{$\mu$} & \multicolumn{1}{c}{$\zeta_{\rm R}$} \\

\cmidrule(lr){1-1} \cmidrule(lr){2-2} \cmidrule(lr){3-6} 

\multirow{1}{*}{$0$}    & $-0.019010(2)$ & $-0.04147$ & $-0.02294$ & $-0.95615$ & $-0.14969$ \\

\multirow{1}{*}{$0.15$} & $0.154215(6)$  & $-0.11783$ & $0.14001$  & $-0.29534$ & $-0.33552$ \\

\hline \hline \\

 & \multicolumn{5}{c}{$16\times16\times3=768$-site cluster} \\

\cmidrule(lr){2-6}
 
 & \multicolumn{1}{c}{$U(1)$} & \multicolumn{4}{c}{$\mathbb{Z}_{2}$} \\

\cmidrule(lr){2-2} \cmidrule(lr){3-6}

\multicolumn{1}{c}{$J_{2}/J_{1}$} & \multicolumn{1}{c}{$\chi_{2}$} & \multicolumn{1}{c}{$\Delta_{2}$} & \multicolumn{1}{c}{$\chi_{2}$} & \multicolumn{1}{c}{$\mu$} & \multicolumn{1}{c}{$\zeta_{\rm R}$} \\ 

\cmidrule(lr){1-1} \cmidrule(lr){2-2} \cmidrule(lr){3-6} 

\multirow{1}{*}{$0$}    & $-0.019522(3)$ & $-0.03646$ & $-0.02103$ & $-0.90133$ & $-0.14296$ \\ 
\multirow{1}{*}{$0.15$} & $0.153695(4)$  & $-0.11826$ & $0.14187$  & $-0.29038$ & $-0.33811$ \\ 

\hline \hline
\end{tabular}
\caption{Optimized parameters of the extended $U(1)$ Dirac SL (current work) and the $\mathbb{Z}_{2}[0,\pi]\beta$ 
state (corresponding to Refs.~\cite{li2016,privcomm}). The NN hopping is $\chi_{1}=1$. The corresponding energies are given in Table~\ref{tab:energies}.}
\label{tab:parameters}
\end{table*}

\end{document}